\documentclass[conference]{IEEEtran}
\usepackage{graphicx}
\usepackage{color}
\usepackage[usenames,dvipsnames,svgnames,table]{xcolor}
\usepackage{boldline}

\usepackage[english]{babel}
\usepackage{blindtext}

\DeclareGraphicsExtensions{.eps}
\usepackage{multirow}

\usepackage{siunitx}
\usepackage{cite}
\ifCLASSINFOpdf
\else
\fi
\usepackage[cmex10]{amsmath}
\usepackage{footnote}
\makesavenoteenv{tabular}
\makesavenoteenv{table}

\usepackage[footnotesize, tight]{subfigure}

\usepackage[top=0.7in, bottom=1.1in, left=0.625in, right=0.625in]{geometry}

\DeclareRobustCommand*{\IEEEauthorrefmark}[1]{%
  \raisebox{0pt}[0pt][0pt]{\textsuperscript{\footnotesize\ensuremath{#1}}}}

\begin{document}
\title{Outdoor to Indoor Penetration Loss at 28 GHz for Fixed Wireless Access}


%
\author{
    \IEEEauthorblockN{C. U. Bas\IEEEauthorrefmark{1}, {\it Student Member, IEEE},
    R. Wang\IEEEauthorrefmark{1}, {\it Student Member, IEEE},
    T. Choi\IEEEauthorrefmark{1}, {\it Student Member, IEEE}, \\
    S. Hur\IEEEauthorrefmark{3}, {\it Member, IEEE},
    K. Whang\IEEEauthorrefmark{3}, {\it Member, IEEE},
    J. Park\IEEEauthorrefmark{3}, {\it Member, IEEE}, \\
    J. Zhang\IEEEauthorrefmark{2}, {\it Fellow, IEEE}, 
    A. F. Molisch\IEEEauthorrefmark{1}, {\it Fellow, IEEE}  
    }\\
    \IEEEauthorblockA{\IEEEauthorrefmark{1}University of Southern California, Los Angeles, CA, USA,}
    
    \IEEEauthorblockA{\IEEEauthorrefmark{2}Samsung Research America, Richardson, TX, USA}
    
     \IEEEauthorblockA{\IEEEauthorrefmark{3}Samsung Electronics, Suwon, Korea}
}

\maketitle
\IEEEpeerreviewmaketitle

\begin{abstract}
  
This paper present the results from a 28 GHz channel sounding campaign performed to investigate the effects of outdoor to indoor penetration on the wireless propagation channel characteristics for an urban microcell in a fixed wireless access scenario. The measurements are performed with a real-time channel sounder, which can measure path loss up to 169 dB, and equipped with phased array antennas that allows electrical beam steering for directionally resolved measurements in dynamic environments. Thanks to the short measurement time and the excellent phase stability of the system, we obtain both directional and omnidirectional channel power delay profiles without any delay uncertainty. For outdoor and indoor receiver locations, we compare path loss, delay spreads and angular spreads obtained for two different types of buildings.

\end{abstract}

\section{Introduction}

The number of connected devices and their data requirements have been increasing exponentially. Especially with the introduction of new technologies such as augmented reality, virtual reality and Ultra-HD video streaming, monthly global IP traffic is expected to reach 278 exabytes by 2021 \cite{forecast2017cisco}. So far this demand for broadband internet is fulfilled mostly with fiber-optical links. An approach challenging costly fiber-optical deployments is leveraging the use of Fixed Wireless Access (FWA), also known as Local Multi-point Distribution Service (LMDS). 

When combined with the large bandwidth available at millimeter-wave (mm-wave) frequencies, Gbps broadband connections to multiple users can be realized via FWA \cite{pi2016millimeter} \cite{wells2009faster}. The prospect of utilizing the fallow spectrum at the frequencies higher than \SI{6}{GHz} fueled the interest in mm-wave propagation channel measurements \cite{papazian1997study} \cite{Molisch_2016_eucap} \cite{rappaport2013millimeter}. Especially, the \SI{28}{GHz} band attracts a lot of interest thanks to comparatively lower hardware and implementation costs due to a relatively lower carrier frequency.

An accurate channel model is crucial for an efficient wireless system design. Outdoor to indoor penetration loss is one of the most important factors of affecting the deployment of FWA or similar systems. Especially at mm-wave frequencies, the penetration loss is shown to be higher and highly sensitive to the small changes in the material types. Furthermore, at mm-wave frequencies, the angular characteristics of the propagation channel are of the utmost importance, since these systems envisioned to be highly dependent on the beam-forming gain to overcome higher path loss at higher frequencies \cite{Roh2014millimeter}.

Most of the current literature at \SI{28}{GHz} focus on outdoor to outdoor measurements. Deviating from this trend, \cite{Ko_2016_feasibility} performed outdoor to indoor measurements at \SI{28}{GHz} by using a rotating horn antenna channel sounder which can provide an accurate absolute delay information. They observed a larger number of clusters, larger excess delays and larger angular spreads indoor. However, there was a limited number of indoor RX locations and a single type of building was investigated. In \cite{larsson_2014_outdoor}, the authors showed that the excess loss due to outdoor to indoor penetration at \SI{28}{GHz} can vary from \SI{3}{dB} to \SI{60}{dB} depending on the RX location and the construction material types. Ref. \cite{rodriguez_2017_empirical} investigates the penetration loss for external and internal walls at different carrier frequencies ranging from \SI{0.8}{GHz} to \SI{28}{GHz}, and proposes a linear model for frequency dependency of the penetration loss. In \cite{zhao_2013_28GHz}, the authors present results for penetration loss and reflection coefficient measurements for different types of building materials. For example; the penetration loss for clear glass is measured as \SI{3.9}{dB} while it is \SI{40}{dB} for tinted glass. A summary of penetration loss results, and a frequency-dependent model, is given in \cite{haneda20165g} and also used in 3GPP. None of references \cite{larsson_2014_outdoor, rodriguez_2017_empirical, zhao_2013_28GHz, haneda20165g}  provide any delay or angular statistics.

There are other studies investigating the outdoor to indoor propagation channel at different mm-wave frequencies. Ref. \cite{tra_2016_outdoor} provides delay and angular statistics at \SI{20}{GHz}. For a urban microcellular environment, \cite{imai_2016_outdoor} presents the penetration loss at \SI{26}{GHz} and \SI{37}{GHz}, and compares these with microwave frequencies. Finally the references \cite{kim_2016_mmwave}, \cite{Cheng_2016_study} and \cite{Diakhate_2017_millimeter} discusses outdoor to indoor propagation measurement campaigns performed around \SI{60}{GHz}.

In this work, we present the results from a \SI{28}{GHz} channel sounding campaign performed to investigate the effects of outdoor to indoor penetration on the channel characteristics for a fixed wireless access scenario. Similar to a real deployment, the measurements capture the combined effect of path loss, foliage loss and the outdoor to indoor penetration loss. A real-time channel sounder equipped with phased array antennas was used for the measurements \cite{bas_2017_realtime}. The phased arrays form beams at the different TX and RX angles and switch between these beams in microseconds, enabling directionally resolved results while ensuring minimal variation in the environment during the measurements. We provide examples of power delay profiles and extracted multi-path components (MPC) and compare path loss, delay spread and angular characteristics for indoor and outdoor RX locations. 

The rest of the paper is organized as follows. Section \ref{sec:setup} describes the measurement equipment and the configuration during this channel sounding campaign. Section \ref{sec:meas} provides details about the two measurement scenarios under investigation. Section \ref{sec:results} presents results for path loss, delay and angular statistics for indoor and outdoor RX locations. Finally, Section \ref{sec:conc} summarizes results and suggests future work.

\section{Measurement Setup}  \label{sec:setup}
  \normalsize
  \begin{table}[tbp]\centering
  \caption{Sounder specifications}
  \renewcommand{\arraystretch}{1.3}
\begin{tabular}{l|c}
    \hline
    \multicolumn{2}{c}{Hardware Specifications} \\ \hline \hline
    Center Frequency & 27.85 GHz\\
    Instantaneous Bandwidth & 400 MHz\\
    Antenna array size & 8 by 2 \\
    Horizontal beam steering (RX/TX)& $360^\circ$ / $90^\circ$\\
    Horizontal 3dB beam width & $12^\circ$\\
    Vertical beam steering (RX/TX) & $-30^\circ$ to $30^\circ$ / 0 \\
    Vertical 3dB beam width & $22^\circ$\\
    Horizontal/Vertical steering steps & $5^\circ$ / $10^\circ$\\
    Beam switching speed & 2$\mu s$ \\
    TX EIRP & 57 dBm \\
    RX noise figure & $\le$ 5 dB \\ 
    ADC/AWG resolution & 10/15-bit \\
    Data streaming speed & 700 MBps \\ \hline
    \multicolumn{2}{c}{Sounding Waveform Specifications} \\ \hline \hline
    Waveform duration & 2 $\mu s$ \\
    Repetition per beam pair & 10 \\
    Number of tones & 801 \\
    Tone spacing & 500 kHz \\
    PAPR & 0.4 dB \\ 
    Total sweep time & 101.08 ms (each $90^\circ$ RX Sector)\\ \hline      
  \end{tabular} \label{specs}
\end{table}

In this campaign, we used a switched-beam, wide-band mm-wave sounder with 400 MHz real-time bandwidth \cite{bas_2017_realtime}.The sounding signal is a multi-tone signal which consists of equally spaced 801 tones covering 400 MHz. A low peak to average power ratio (PAPR) of $0.4$ dB is achieved by manipulating the phases of individual tones as suggested in \cite{Friese1997multitone}. This allows us to transmit with power as close as possible to the 1 dB compression point of the power amplifiers without driving them into saturation.  

Both the TX and the RX are equipped with 2 by 8 rectangular phased array antennas capable of forming beams that can be electronically steered with $5^{\circ}$ resolution in the range of $[-45^{\circ}, 45^{\circ}]$ in azimuth and $[-30^{\circ}, 30^{\circ}]$ in elevation. During this measurement campaign we utilize a single elevation angle $0^{\circ}$ with 19 azimuth angles for the TX, and 7 elevation angles along with 19 azimuth angles for the RX. With an averaging factor of 10, the total sweep time is \SI{101.08}{ms} for 2527 total beam pairs. Since phased arrays cover $90^{\circ}$ sectors, we rotated the RX to $\{0^{\circ},90^{\circ},180^{\circ},270^{\circ}\}$ to cover $360^{\circ}$ while using a single orientation at the TX. Consequently, for each measurement location, we obtain a frequency response matrix of size 7 by 72 by 19 by 801.  

Moreover, thanks to the beam-forming gain, the TX EIRP is \SI{57}{dBm}, and the measurable path loss is \SI{159}{dB} without considering any averaging or spreading gain. Including the averaging ratio used in this campaign the measurable path-loss is \SI{169}{dB}. By using GPS-disciplined Rubidium frequency references, we were able to achieve both short-time and long-time phase stability. Combined with the short measurement time this limits the phase drift between TX and RX, enabling phase-coherent sounding of all beam pairs even when TX and RX are physically separated and have no cabled connection for synchronization. Consequently, the directional power delay profiles (PDP) can be combined easily to acquire the omnidirectional PDP. Table \ref{specs} summarizes the detailed specification of the sounder and the sounding waveform. References \cite{bas_2017_realtime} and \cite{bas_realjournal_2017} discuss further details of the sounder setup, the validation measurements, and the data processing.

\section{Measurement Campaign} \label{sec:meas}

The measurements were performed at two different locations on the University of Southern California campus. In both cases, the RX was on the first floor, while the TX height was on a scissor lift at the height of \SI{5}{m} imitating an urban micro-cell (UMi) scenario. To investigate the more challenging cases, we chose buildings surrounded by foliage and made sure the angle of the direct path is narrow with respect to the front facade of the target building.

\begin{figure}[tbp]
        \centering\includegraphics[width=0.9\linewidth]{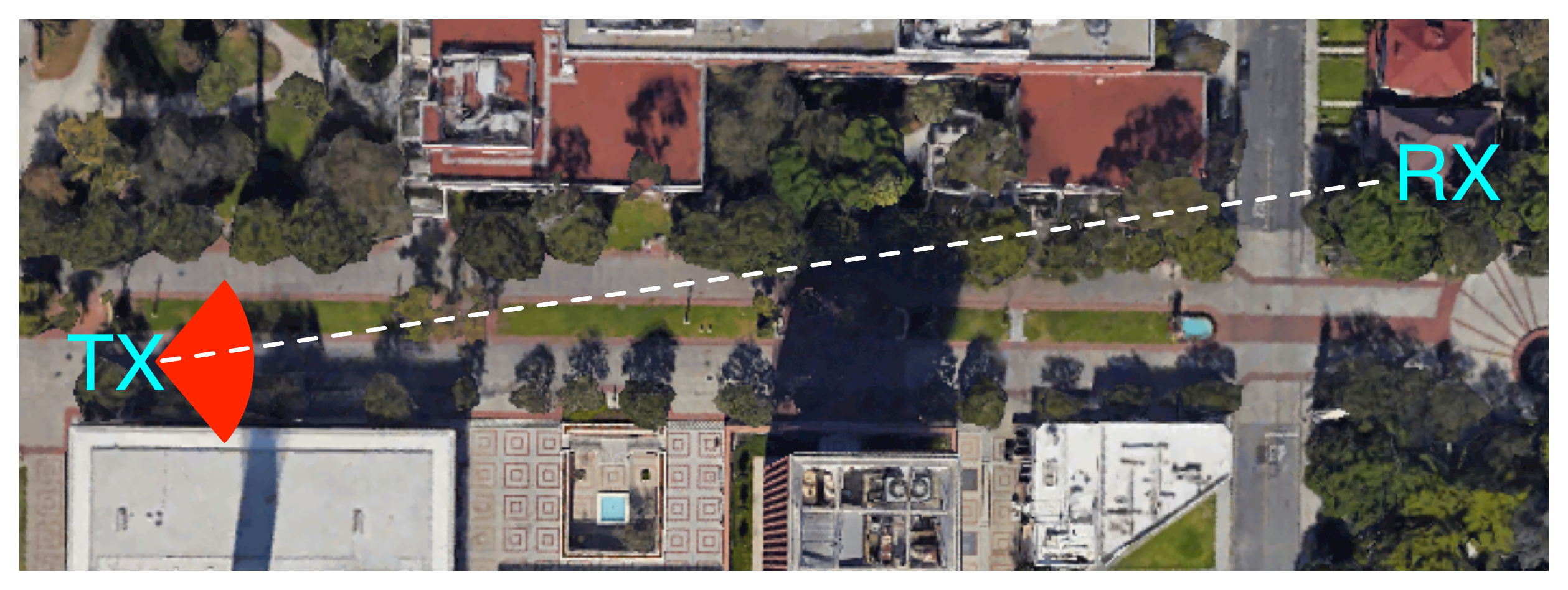}\caption{TX and RX locations for SFU}\label{fig:JEP_outside}
\end{figure}

\begin{figure}[tbp]
        \centering\includegraphics[width=0.9\linewidth]{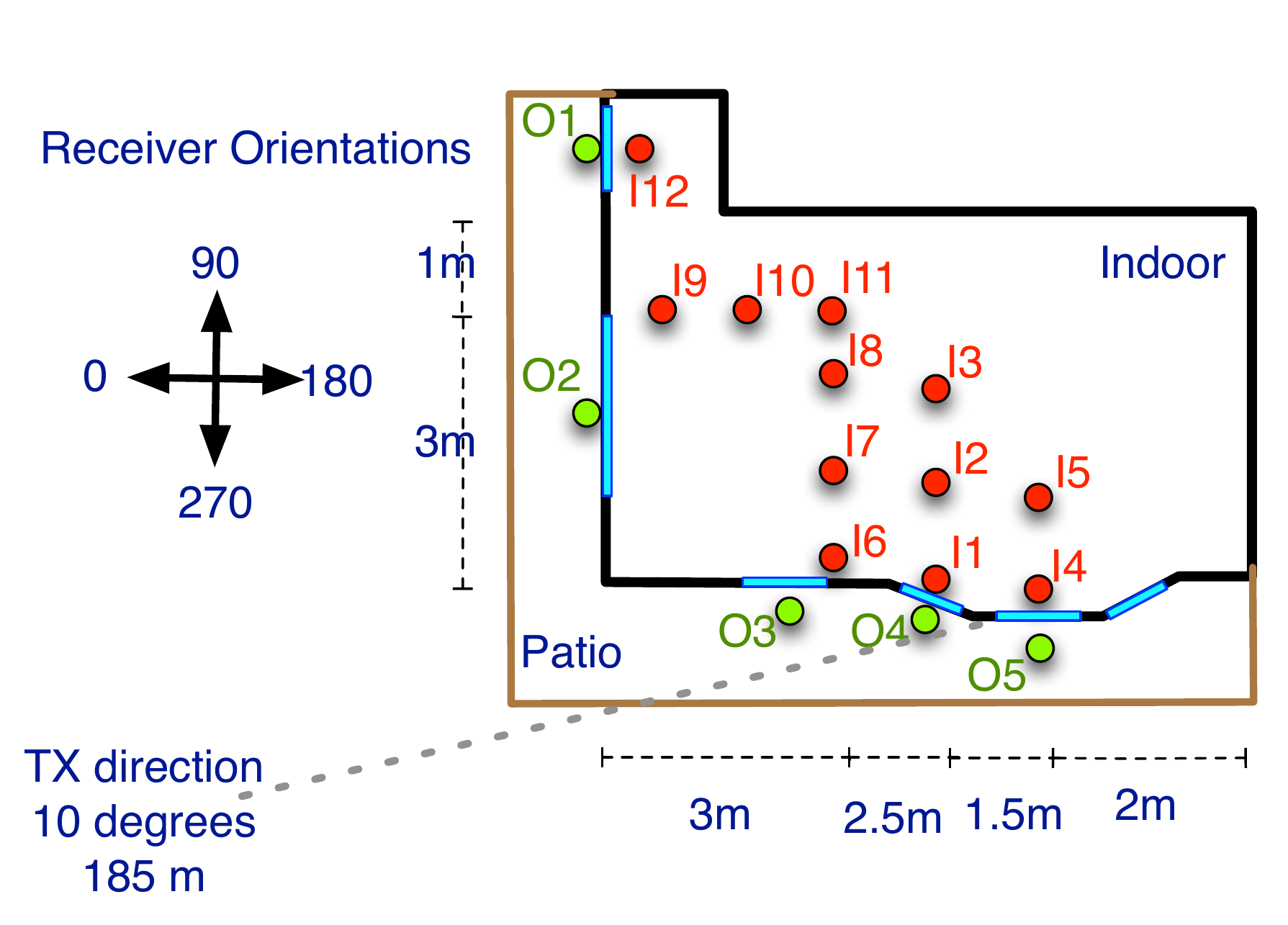}\caption{Layout of receivers for SFU}\label{fig:JEP_layout}
\end{figure}

\subsection{Single Family Unit}

Figure \ref{fig:JEP_outside} shows TX and RX locations on the campus for the first location. The building is a two-story, detached, single family unit using wood frames, as is typical in California. . There is also a covered, first-floor patio wrapping around the building. Measurement points are marked in Figure \ref{fig:JEP_layout}. There are 5 outdoor points placed right in front of the windows on the patio. 12 indoor measurement locations are placed throughout the room as the furniture allows. All RX points have the same height. The TX is placed on the same street with the house at a distance of \SI{185}{m} at an angle of $10^\circ$ according to the given directions in Figure \ref{fig:JEP_outside}. The direct path from TX to the house is blocked by foliage from trees. Additionally, points {\em O1} and {\em O2} are also shadowed by the building across the street.

\subsection{Multi-Story Building}
\begin{figure}[tbp]
        \centering\includegraphics[width=0.9\linewidth]{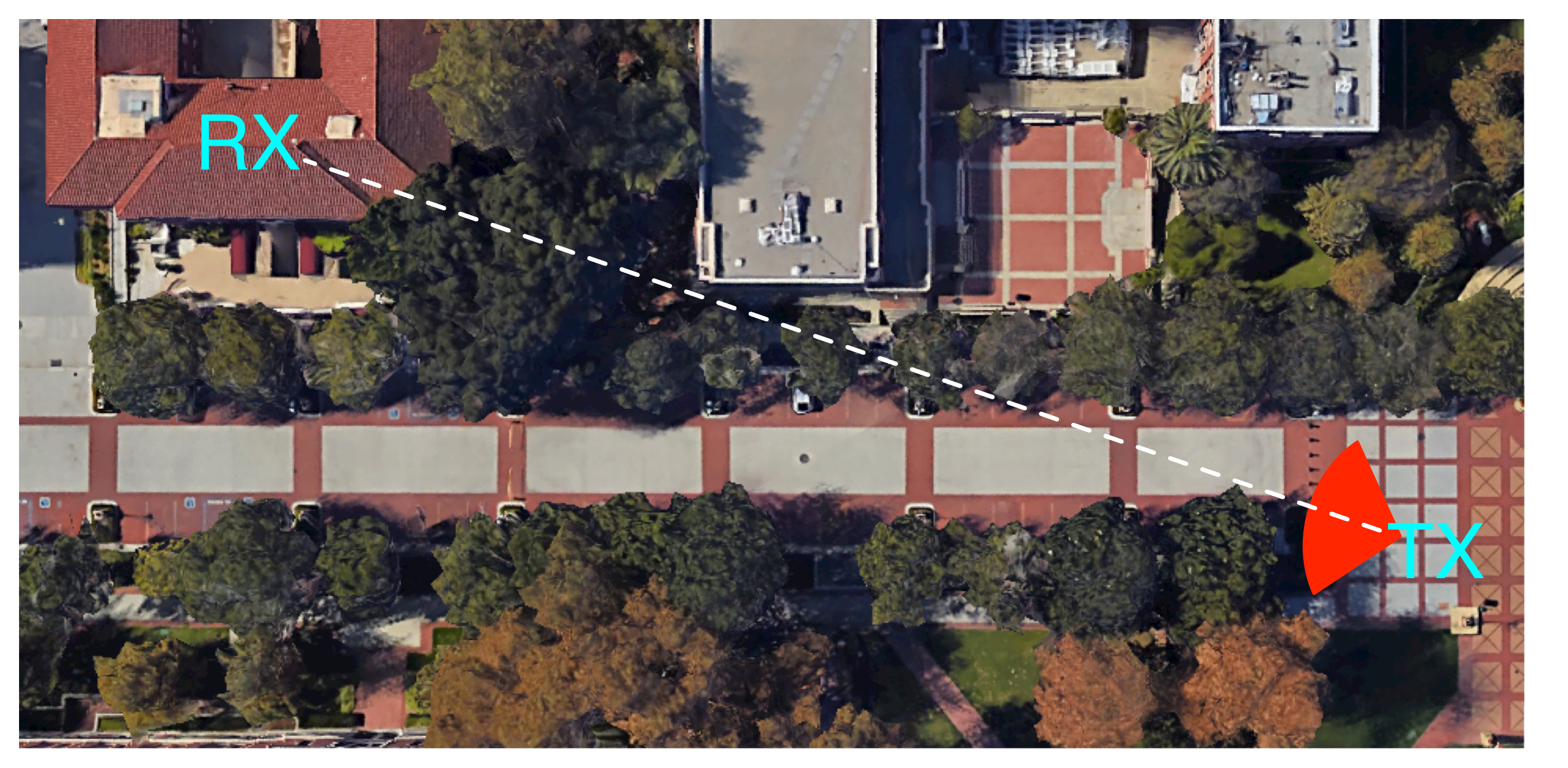}\caption{ TX and RX locations for MSB}\label{fig:JHH_outside}
\end{figure}

\begin{figure}[tbp]
        \centering\includegraphics[width=0.9\linewidth]{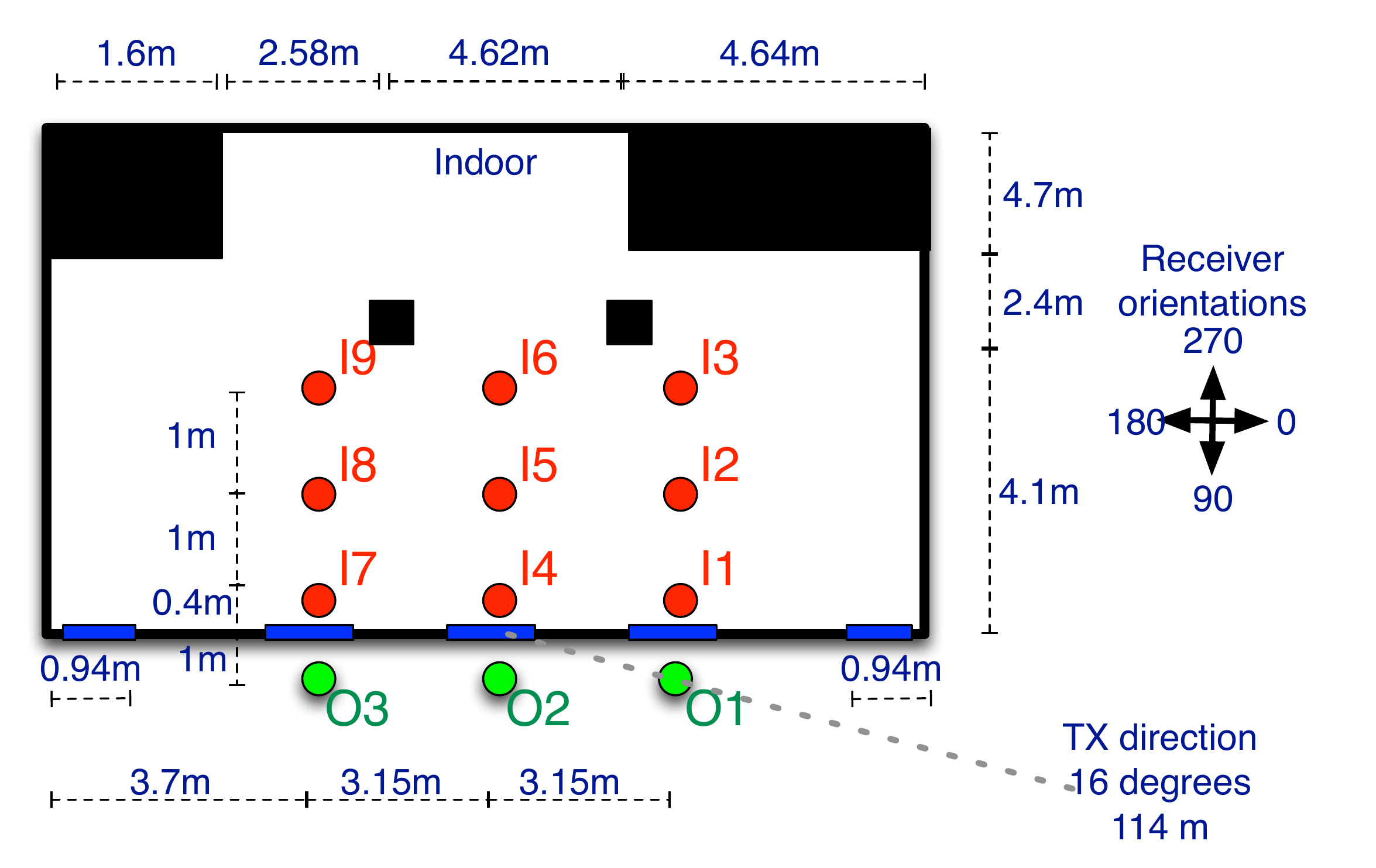}\caption{Layout of receivers in the MSB}\label{fig:JHH_layout}
\end{figure}

The second building is a multi-story, brick building (MSB) surrounded by heavy foliage as shown in Figure \ref{fig:JHH_outside}. The TX-RX distance is \SI{114}{m}, and the angle of the direct path is $16^\circ$. Three outdoor measurement locations are just outside of the three front-facing windows. For each window, there are three indoor locations placed on a line along with the corresponding outdoor point. The distances from the indoor measurement points to the window are \SI{40}{cm}, \SI{1.4}{m} and \SI{2.4}{m}, Figure \ref{fig:JHH_layout}.

\begin{figure}[tbp]
        \centering\includegraphics[width=0.9\linewidth]{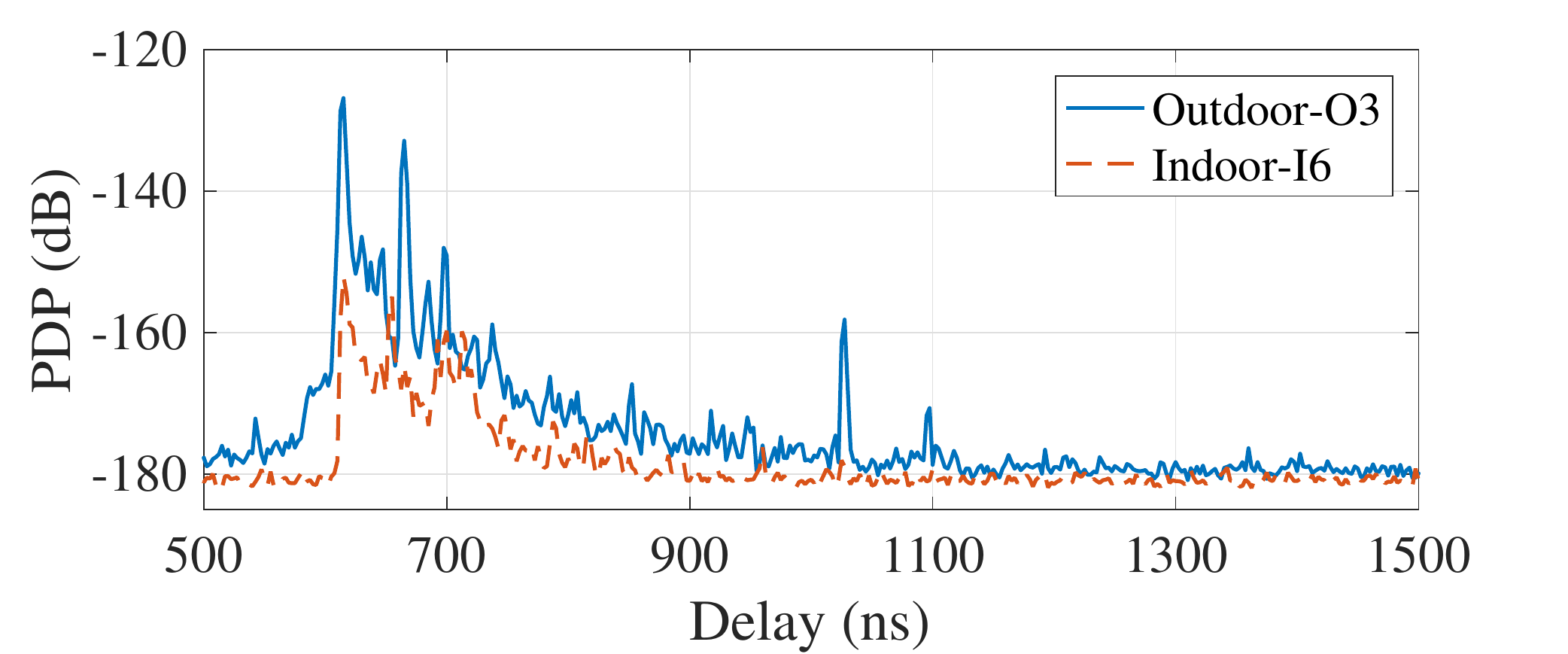}\caption{Power delay profiles for sample indoor and outdoor points}\label{fig:JEP_pdp}
\end{figure}

\begin{figure}[tbp]
        \centering\includegraphics[width=0.7\linewidth, viewport=70 40 550 400, clip]{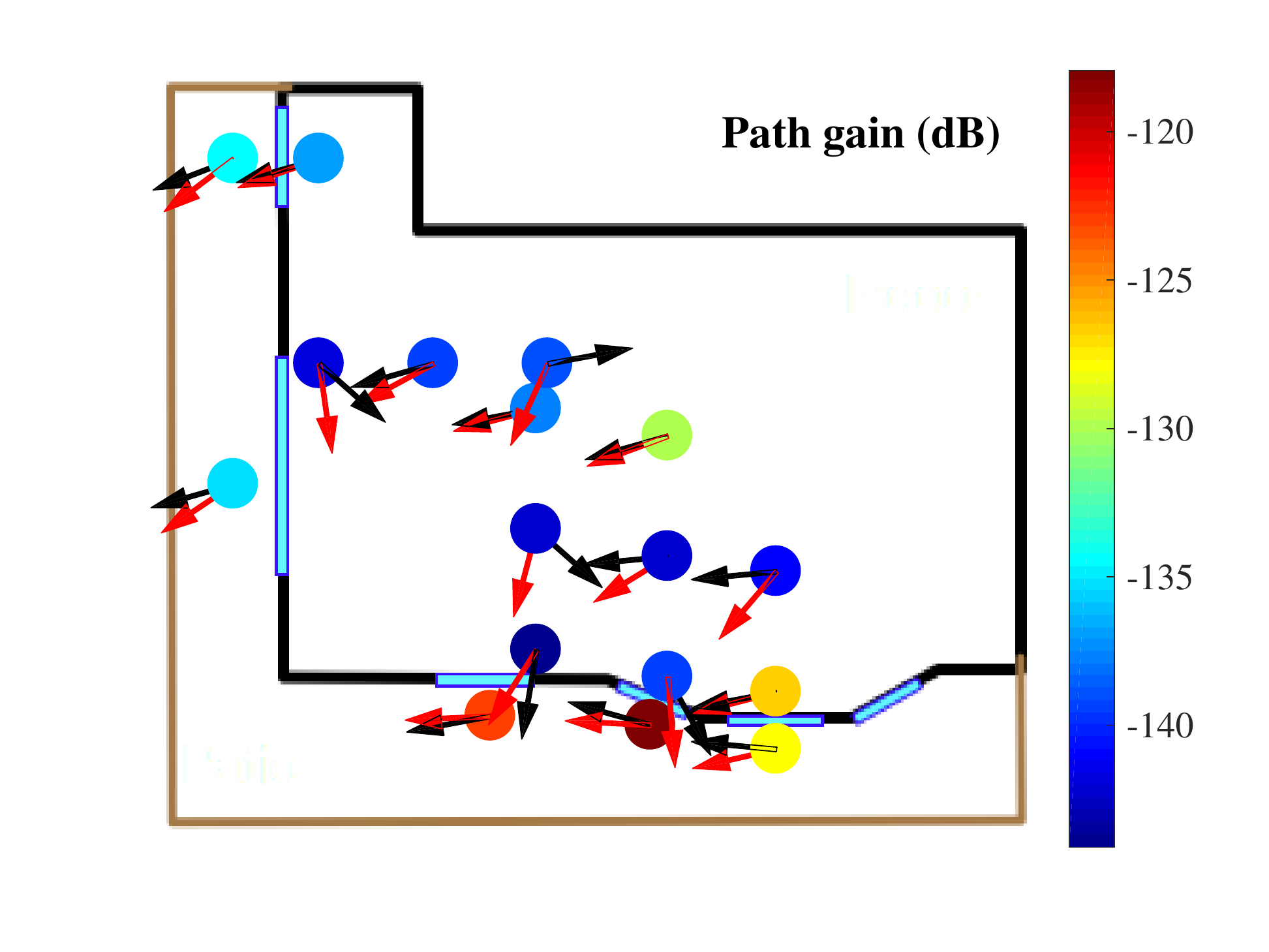}\caption{Path gains in dB for the RX locations in the SFU, red arrows indicate the mean direction of arrivals, black arrows indicate the RX beam direction with the highest power. }\label{fig:JEP_pg}
\end{figure}

 \begin{figure}[tbp]
        \centering\includegraphics[width=0.9\linewidth]{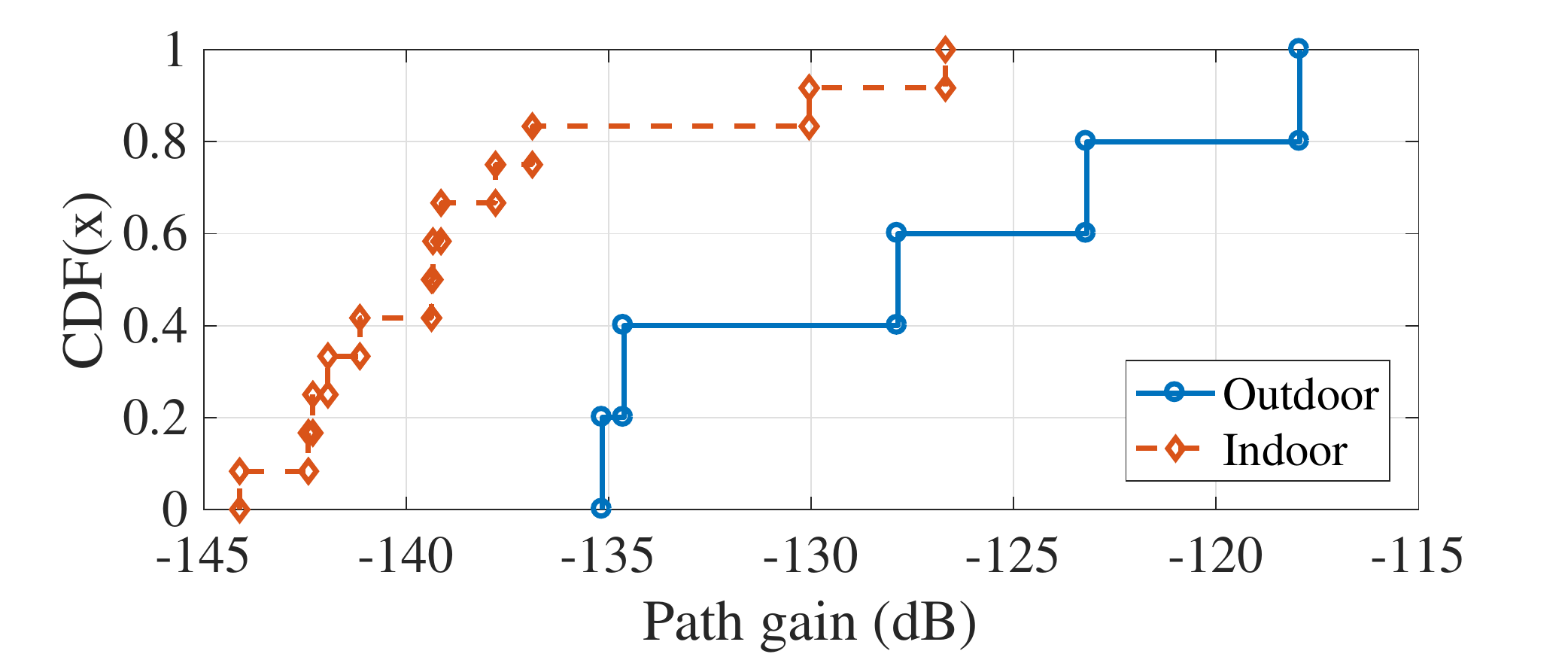}\caption{CDF of path gains in the SFU}\label{fig:JEP_pg_cdf}
\end{figure}

\begin{figure}[tbp]
        \centering\includegraphics[width=0.9\linewidth]{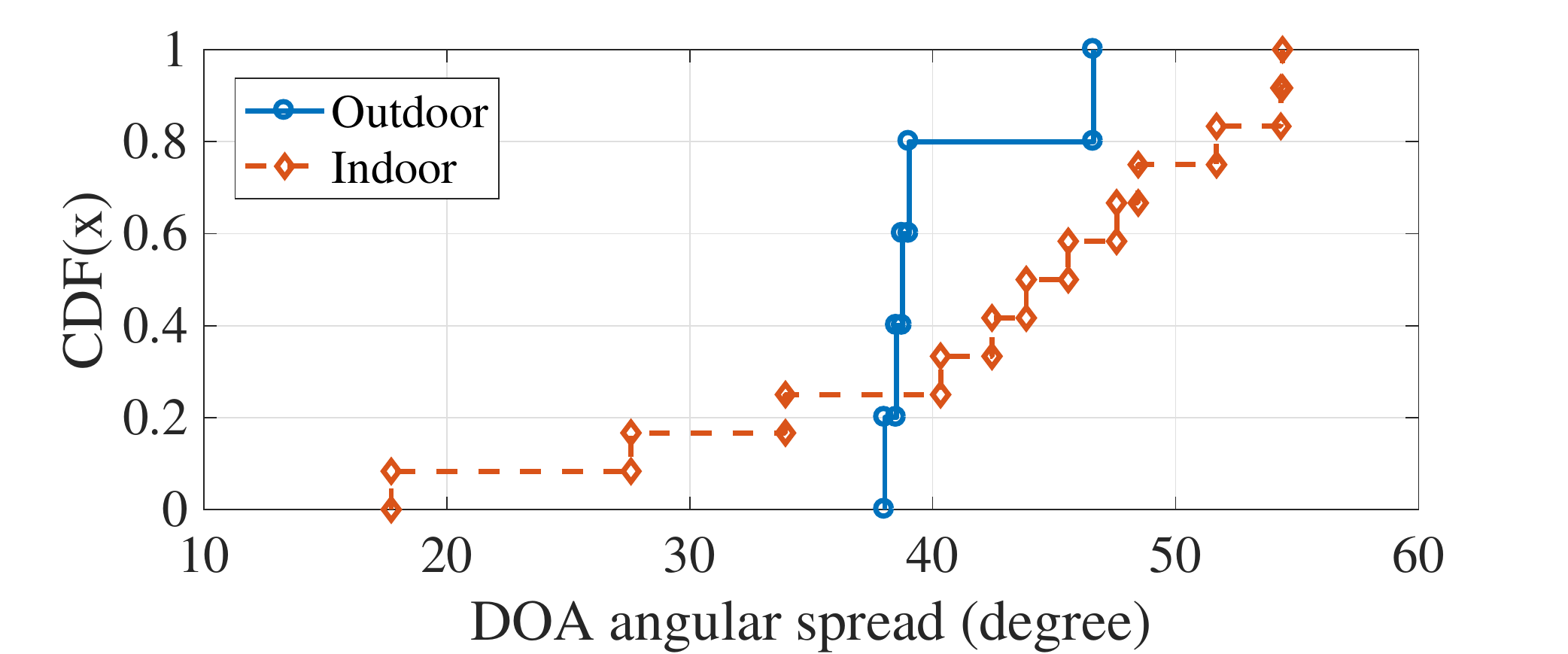}\caption{CDF of DoA angular spread in the SFU}\label{fig:JEP_as_cdf}
\end{figure}

\section{Results}\label{sec:results}

In this section we compare the path-loss, root mean square delay spreads (RMS-DS) and direction of arrival (DoA) statistics for outdoor and indoor measurements for both scenarios.

\subsection{Single Family Unit}

Figure \ref{fig:JEP_pdp} shows the power delay profiles (PDP) for an outdoor ({\em O3}) and an indoor ({\em I6}) RX points. In this particular case, the excess loss due to penetration is as high as \SI{21}{dB}. 

Figures \ref{fig:JEP_pg} and \ref{fig:JEP_pg_cdf} respectively show the path gain values and their cumulative distribution functions for all RX points. The free space path loss for \SI{185}{m} at \SI{28}{GHz} is \SI{106.7}{dB}. However due shadowing from foliage, the path-loss for the outdoor RX locations varies between \SI{117}{dB} to \SI{135}{dB} with a mean of \SI{127.8}{dB}. According to the path-loss model in \cite{bas_2017_microcell}, based on measurements in a similar environment, the anticipated path loss is \SI{127.4}{dB} for the distance of \SI{185}{m} which shows a good agreement with our results. The mean path loss for indoor locations is \SI{138.4}{dB}, which is \SI{10.6}{dB} higher than the mean-outdoor path loss. However if the immediate points outside and inside of the windows are compared, the excess loss vary from \SI{1}{dB} to \SI{21.5}{dB} depending on the windows' positions. 

The mean (DoA) and the directions of the RX beam with the highest power in azimuth are also shown in Figure \ref{fig:JEP_pg}. Although the mean angular spreads are similar for outdoor and indoor measurements, the ranges of values differ significantly as seen in Figure \ref{fig:JEP_as_cdf}.  This has important implications for system design. Deployment of antennas indoors might require higher adaptivity of the antenna pattern, and the angular spread could change drastically when relocating the antenna within a room. 

\begin{figure}[tbp]
        \centering\includegraphics[width=0.7\linewidth, viewport=70 40 550 400, clip]{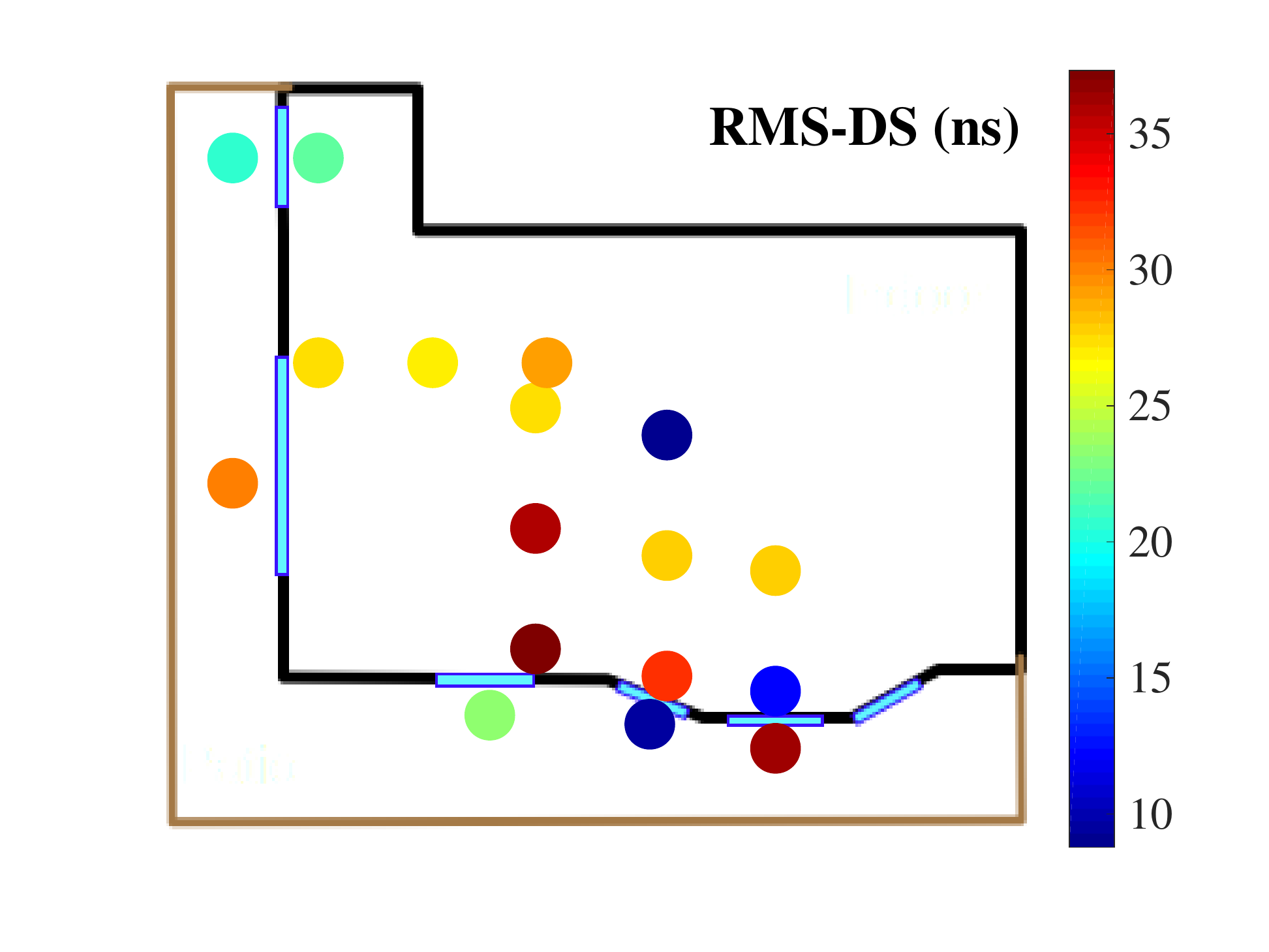}\caption{RMS-DS in ns for the RX locations in the SFU}\label{fig:JEP_rmsds}
\end{figure}

\begin{figure}[tbp]
        \centering\includegraphics[width=0.9\linewidth]{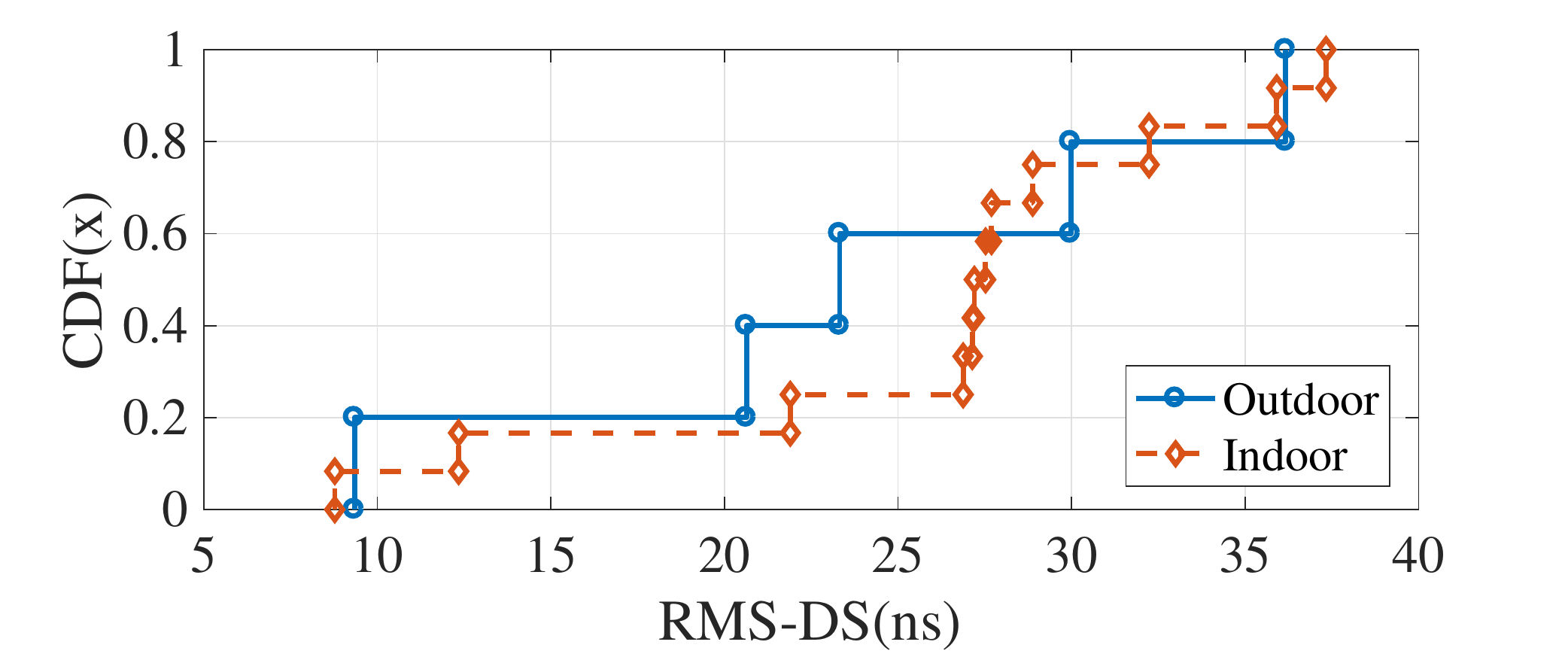}\caption{CDF of RMS-DS in the SFU}\label{fig:JEP_rmsds_cdf}
\end{figure}

Figure \ref{fig:JEP_rmsds} shows the RMS-DS. As also seen in the CDFs in Figure \ref{fig:JEP_rmsds_cdf}, indoor and outdoor RMS-DS values do not differ significantly. As summarized in Table \ref{tab:rmsds} the means, the medians, and the ranges of RMS-DS are fairly similar for the two cases. Furthermore, when the two data sets are compared by using a two-sample Kolmogorov-Smirnov test \cite{massey_1951_kolmogorov}, the null hypothesis of these two data coming from the same underlying distribution is not rejected with a 5\% significance level. 

\begin{table}[h]  
\centering \normalsize \caption{RMS-DS (ns) for SFU}
  \begin{tabular} {l|c|c|c|c}
    & Mean & Median & Max & Min \\ \hline
    Outdoor &  23.89 &  23.32 & 36.17  &  9.34 \\
    Indoor & 26.15 & 27.36 & 37.33 &   8.77  \\
  \end{tabular}\label{tab:rmsds}
\end{table}

\subsection{Multi-Story Building}

\begin{figure}[tbp]
        \centering\includegraphics[width=0.7\linewidth, viewport=70 40 550 400, clip]{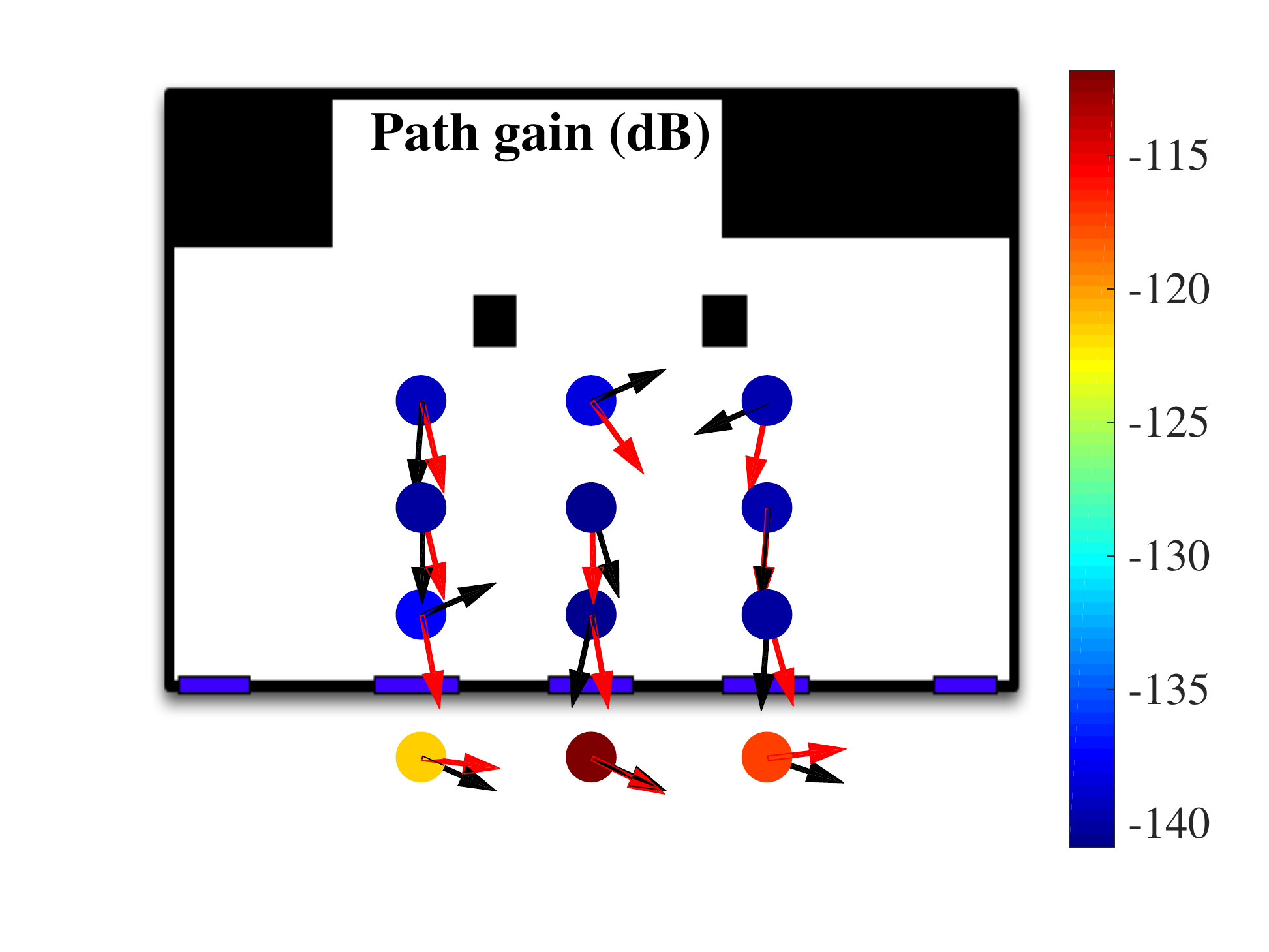}\caption{Path gains in dB for the RX locations in MSB, red arrows indicate mean direction of arrivals, , black arrows indicate the RX beam direction with the highest power.}\label{fig:JHH_pg}
\end{figure}

Figure \ref{fig:JHH_pg} shows the path gains for all measurement locations in the MSB. Since for all indoor locations, multi-path components undergo similar propagation paths, the observed path gains don't vary significantly. The mean path losses for outdoor and indoor locations are \SI{117}{dB} and \SI{139.7}{dB}, respectively. Hence mean excess loss due to outdoor to indoor penetration is \SI{22.7}{dB}. As anticipated, due to the brick walls and smaller number of windows compared to the SFU case, the penetration loss is much higher. Note that neither of these buildings uses energy-saving windows (which would have a much higher attenuation).

\begin{figure}[tbp]
        \centering\includegraphics[width=0.9\linewidth]{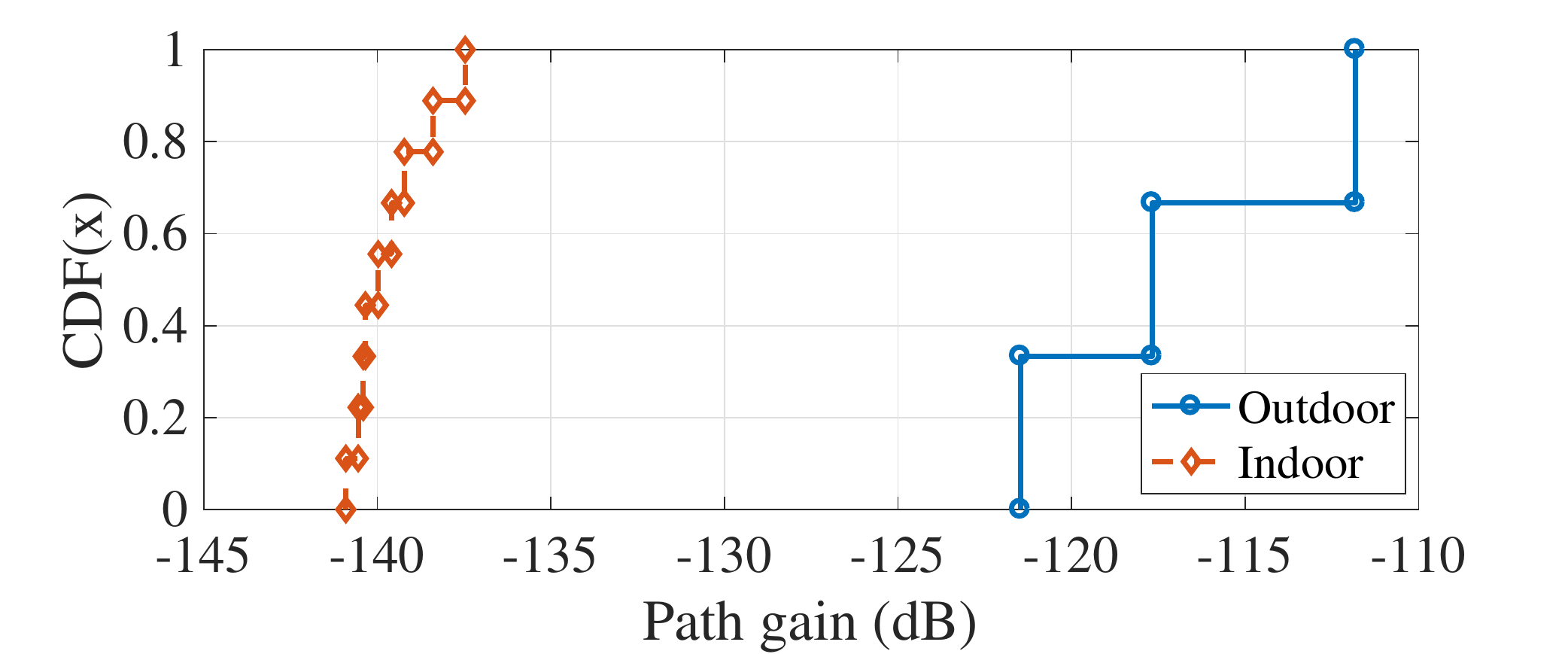}\caption{CDF of path gains in MSB}\label{fig:JHH_pg_cdf}
\end{figure}

\begin{figure}[tbp]
        \centering\includegraphics[width=0.9\linewidth]{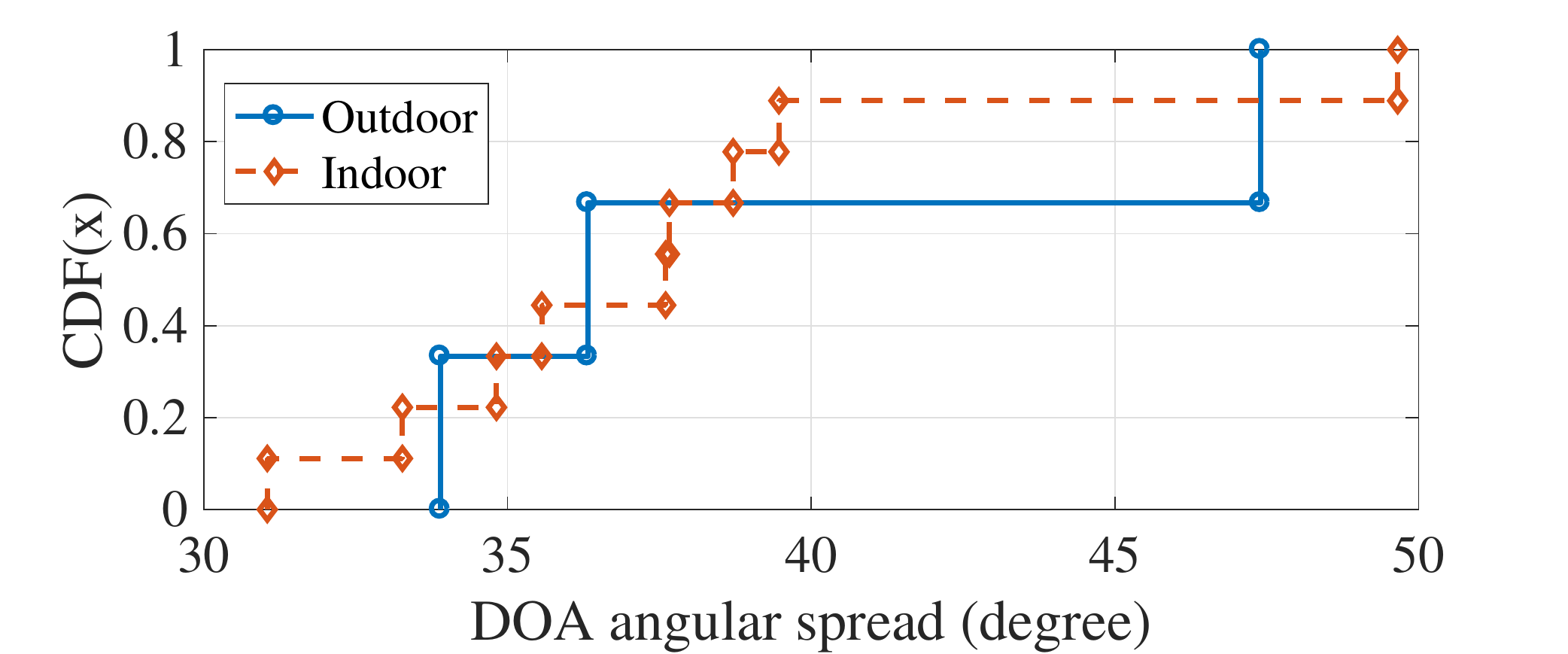}\caption{CDF of of DoA angular spread in MSB}\label{fig:JHH_as_cdf}
\end{figure}

\begin{figure*}[tbp]
        \centering\includegraphics[width=0.9\linewidth,viewport=100 0 1090 260, clip]{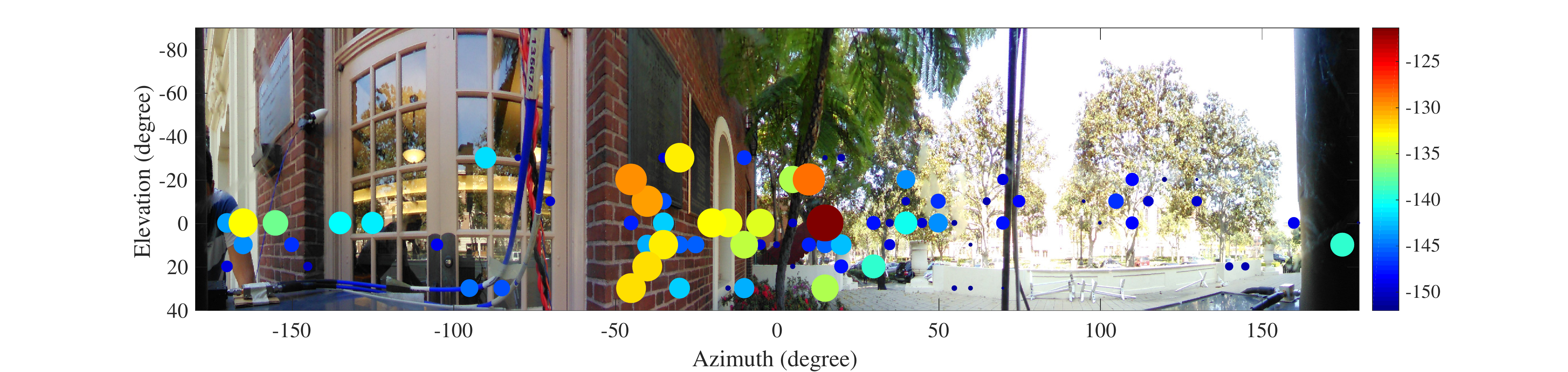}\caption{Detected multipath components vs DoA - outdoor {\em O1} }\label{fig:JHH_o1_angles}
\end{figure*}

\begin{figure*}[tbp]
        \centering\includegraphics[width=0.9\linewidth,viewport=100 0 1090 260, clip]{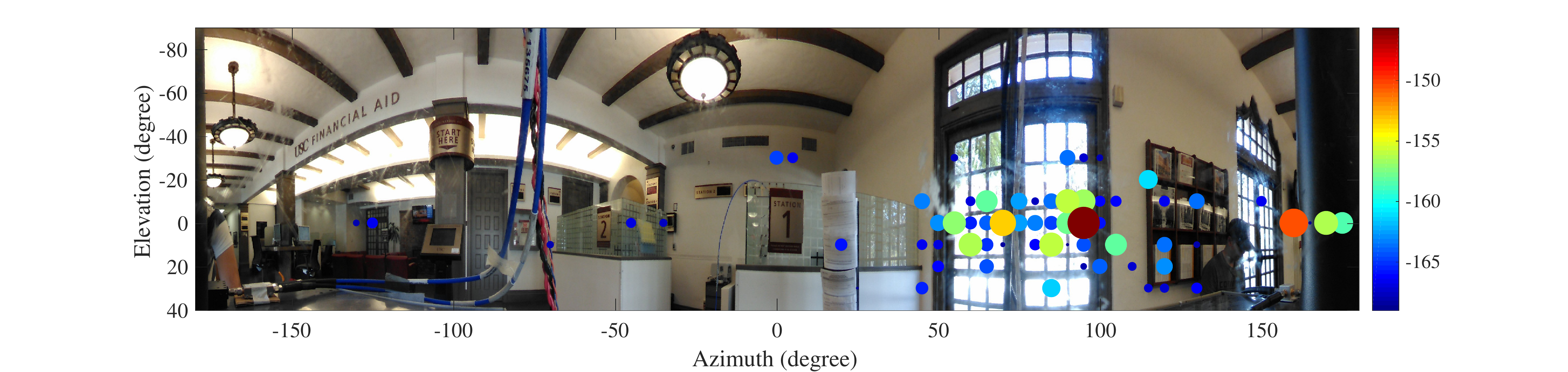}\caption{Detected multi-path components vs DoA - indoor {\em I1}}\label{fig:JHH_i2_angles}
\end{figure*}

\begin{figure}[tbp]
        \centering\includegraphics[width=0.7\linewidth, viewport=70 40 550 400, clip]{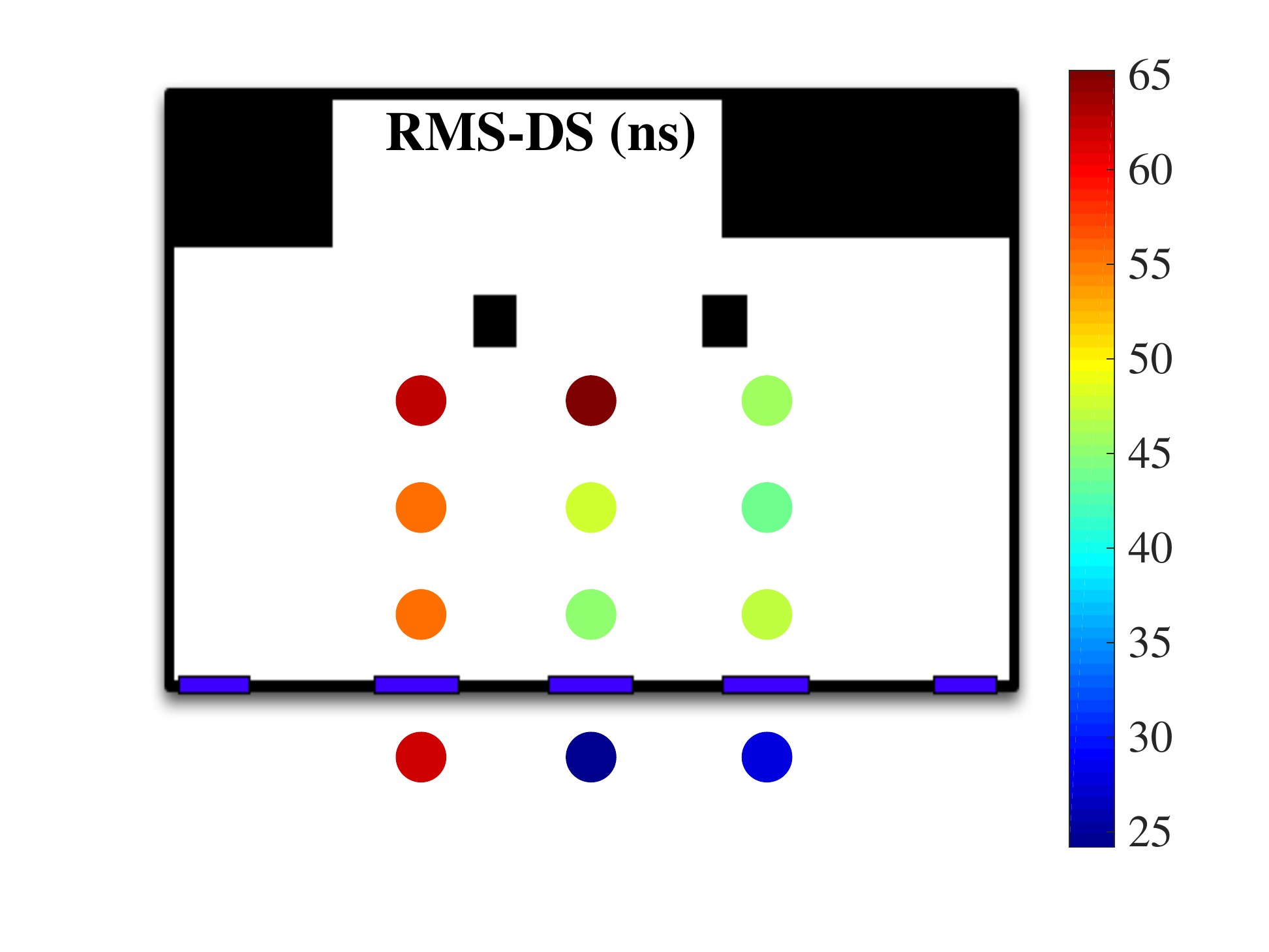}\caption{RMS-DS in ns for the RX locations in the MSB}\label{fig:JHH_rmsds}
\end{figure}

\begin{figure}[htbp]
        \centering\includegraphics[width=0.9\linewidth]{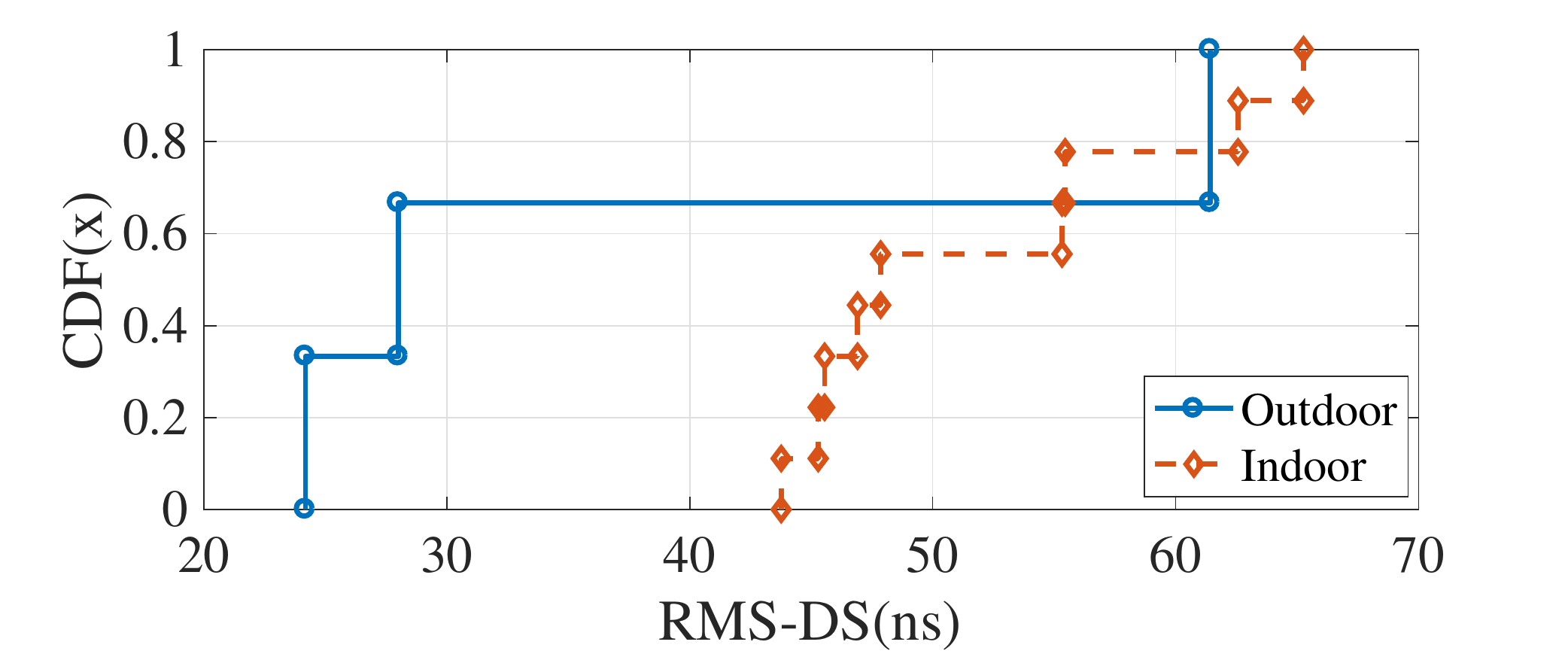}\caption{CDF of RMS-DS in MSB}\label{fig:JHH_rmsds_cdf}
\end{figure}

The detected  MPCs, by using the method described in \cite{bas_realjournal_2017}, are shown in Figures \ref{fig:JHH_o1_angles} and \ref{fig:JHH_i2_angles} for points {\em O1} and {\em I1}, respectively. For all indoor locations except {\em I6} and {\em I7}, we observed similar angular characteristics where all the significant MPCs are through the front windows. For those two points, the strongest MPC was the one through the hallway on the upper right side of the Figure \ref{fig:JHH_pg}. The mean angles and the direction of the best RX beams are marked with red and black arrows in Figure \ref{fig:JHH_pg}. The observed angular spreads are similar for the indoor and outdoor RX points, however, due to the limited number of outdoor locations, it is not certain that the results can be generalized. Compared to the SFU case, the observed angular spread values are within the same range. Figure \ref{fig:JHH_rmsds} shows the RMS-DS. For indoor locations, we observed higher RMS-DS in MSB. In fact, the minimum RMS-DS observed at the MSB-indoor is larger than the maximum RMS-DS for SFU-indoor.

\section{Conclusion} \label{sec:conc}

In this paper, we presented results from a channel sounding campaign focusing on outdoor to indoor wireless propagation channel for a micro-cellular scenario at 28 GHz. We presented results for path-loss, penetration loss, delay spread and angular spread statistics for two different type of housing. We observed mean excess losses of \SI{10.6}{dB} for the single family unit and \SI{22.7}{dB} for the multi-story brick building. For the single-family unit, we also observed that the delay spread statistics are similar for outdoor and indoor receiver locations. However, compared to the single-family unit, the delay spreads in the multi-story building were much larger. In the future, we will perform more measurements to consider other types of buildings and different receiver heights. Finally, we will also try processing the data by using high-resolution parameter extraction such as RIMAX to gain more insights.

\section*{Acknowledgements}
Part of this work was supported by grants from the National Science Foundation and National Institute of Standards and Technology. The authors would like to thank Dimitris Psychoudakis, Thomas Henige, Robert Monroe for their contribution in the development of the channel sounder.

\bibliographystyle{IEEEtran}
\bibliography{mmwave,dynamic,O2I}

\begin{thebibliography}{10}
\providecommand{\url}[1]{#1}
\csname url@samestyle\endcsname
\providecommand{\newblock}{\relax}
\providecommand{\bibinfo}[2]{#2}
\providecommand{\BIBentrySTDinterwordspacing}{\spaceskip=0pt\relax}
\providecommand{\BIBentryALTinterwordstretchfactor}{4}
\providecommand{\BIBentryALTinterwordspacing}{\spaceskip=\fontdimen2\font plus
\BIBentryALTinterwordstretchfactor\fontdimen3\font minus
  \fontdimen4\font\relax}
\providecommand{\BIBforeignlanguage}[2]{{%
\expandafter\ifx\csname l@#1\endcsname\relax
\typeout{** WARNING: IEEEtran.bst: No hyphenation pattern has been}%
\typeout{** loaded for the language `#1'. Using the pattern for}%
\typeout{** the default language instead.}%
\else
\language=\csname l@#1\endcsname
\fi
#2}}
\providecommand{\BIBdecl}{\relax}
\BIBdecl

\bibitem{forecast2017cisco}
C.~V. Forecast, ``Cisco visual networking index: Global mobile data traffic
  forecast update, 2016--2021 white paper,'' \emph{Cisco Public Information},
  2017.

\bibitem{pi2016millimeter}
Z.~Pi, J.~Choi, and R.~Heath, ``Millimeter-wave gigabit broadband evolution
  toward 5g: fixed access and backhaul,'' \emph{IEEE Communications Magazine},
  vol.~54, no.~4, pp. 138--144, 2016.

\bibitem{wells2009faster}
J.~Wells, ``Faster than fiber: The future of multi-g/s wireless,'' \emph{IEEE
  microwave magazine}, vol.~10, no.~3, 2009.

\bibitem{papazian1997study}
P.~B. Papazian, G.~A. Hufford, R.~J. Achatz, and R.~Hoffman, ``Study of the
  local multipoint distribution service radio channel,'' \emph{IEEE
  Transactions on Broadcasting}, vol.~43, no.~2, pp. 175--184, 1997.

\bibitem{Molisch_2016_eucap}
A.~F. Molisch, A.~Karttunen, R.~Wang, C.~U. Bas, S.~Hur, J.~Park, and J.~Zhang,
  ``Millimeter-wave channels in urban environments,'' in \emph{2016 10th
  European Conference on Antennas and Propagation (EuCAP)}, April 2016, pp.
  1--5.

\bibitem{rappaport2013millimeter}
T.~Rappaport, S.~Sun, R.~Mayzus, H.~Zhao, Y.~Azar, K.~Wang, G.~N. Wong, J.~K.
  Schulz, M.~Samimi, and F.~Gutierrez, ``Millimeter wave mobile communications
  for 5g cellular: It will work!'' \emph{Access, IEEE}, vol.~1, pp. 335--349,
  2013.

\bibitem{Roh2014millimeter}
W.~Roh, J.~Y. Seol, J.~Park, B.~Lee, J.~Lee, Y.~Kim, J.~Cho, K.~Cheun, and
  F.~Aryanfar, ``Millimeter-wave beamforming as an enabling technology for 5g
  cellular communications: theoretical feasibility and prototype results,''
  \emph{IEEE Communications Magazine}, vol.~52, no.~2, pp. 106--113, February
  2014.

\bibitem{Ko_2016_feasibility}
J.~Ko, K.~Lee, Y.~J. Cho, S.~Oh, S.~Hur, N.~G. Kang, J.~Park, D.~J. Park, and
  D.~H. Cho, ``Feasibility study and spatial-temporal characteristics analysis
  for 28 ghz outdoor wireless channel modelling,'' \emph{IET Communications},
  vol.~10, no.~17, pp. 2352--2362, 2016.

\bibitem{larsson_2014_outdoor}
C.~Larsson, F.~Harrysson, B.~E. Olsson, and J.~E. Berg, ``An outdoor-to-indoor
  propagation scenario at 28 ghz,'' in \emph{The 8th European Conference on
  Antennas and Propagation (EuCAP 2014)}, April 2014, pp. 3301--3304.

\bibitem{rodriguez_2017_empirical}
I.~Rodriguez, H.~C. Nguyen, I.~Z. Kovács, T.~B. Sørensen, and P.~Mogensen,
  ``An empirical outdoor-to-indoor path loss model from below 6 ghz to cm-wave
  frequency bands,'' \emph{IEEE Antennas and Wireless Propagation Letters},
  vol.~16, pp. 1329--1332, 2017.

\bibitem{zhao_2013_28GHz}
H.~Zhao, R.~Mayzus, S.~Sun, M.~Samimi, J.~K. Schulz, Y.~Azar, K.~Wang, G.~N.
  Wong, F.~Gutierrez, and T.~S. Rappaport, ``28 ghz millimeter wave cellular
  communication measurements for reflection and penetration loss in and around
  buildings in new york city,'' in \emph{2013 IEEE International Conference on
  Communications (ICC)}, June 2013, pp. 5163--5167.

\bibitem{haneda20165g}
K.~Haneda, J.~Zhang, L.~Tan, G.~Liu, Y.~Zheng, H.~Asplund, J.~Li, Y.~Wang,
  D.~Steer, C.~Li \emph{et~al.}, ``5g 3gpp-like channel models for outdoor
  urban microcellular and macrocellular environments,'' in \emph{Vehicular
  Technology Conference (VTC Spring), 2016 IEEE 83rd}.\hskip 1em plus 0.5em
  minus 0.4em\relax IEEE, 2016, pp. 1--7.

\bibitem{tra_2016_outdoor}
N.~Tran, T.~Imai, and Y.~Okumura, ``Outdoor-to-indoor channel characteristics
  at 20 ghz,'' in \emph{2016 International Symposium on Antennas and
  Propagation (ISAP)}, Oct 2016, pp. 612--613.

\bibitem{imai_2016_outdoor}
T.~Imai, K.~Kitao, N.~Tran, N.~Omaki, Y.~Okumura, and K.~Nishimori,
  ``Outdoor-to-indoor path loss modeling for 0.8 to 37 ghz band,'' in
  \emph{2016 10th European Conference on Antennas and Propagation (EuCAP)},
  April 2016, pp. 1--4.

\bibitem{kim_2016_mmwave}
M.~Kim, T.~Iwata, K.~Umeki, K.~Wangchuk, J.~i.~Takada, and S.~Sasaki, ``Mm-wave
  outdoor-to-indoor channel measurement in an open square smallcell scenario,''
  in \emph{2016 International Symposium on Antennas and Propagation (ISAP)},
  Oct 2016, pp. 614--615.

\bibitem{Cheng_2016_study}
\BIBentryALTinterwordspacing
L.~Cheng, W.~Bao, N.~Liu, G.~Yue, X.~Zou, and R.~C. Qiu, ``Study of propagation
  characteristics of outdoor-to-indoor channel in the 60-ghz band,''
  \emph{Journal of Communications and Information Networks}, vol.~1, no.~2, pp.
  93--101, Aug 2016. [Online]. Available:
  \url{https://doi.org/10.1007/BF03391561}
\BIBentrySTDinterwordspacing

\bibitem{Diakhate_2017_millimeter}
C.~A.~L. Diakhate, J.~M. Conrat, J.~C. Cousin, and A.~Sibille,
  ``Millimeter-wave outdoor-to-indoor channel measurements at 3, 10, 17 and 60
  ghz,'' in \emph{2017 11th European Conference on Antennas and Propagation
  (EUCAP)}, March 2017, pp. 1798--1802.

\bibitem{bas_2017_realtime}
C.~U. Bas, R.~Wang, D.~Psychoudakis, T.~Henige, R.~Monroe, J.~Park, J.~Zhang,
  and A.~F. Molisch, ``{A Real-Time Millimeter-Wave Phased Array MIMO Channel
  Sounder},'' in \emph{Vehicular Technology Conference, 2017. VTC 2017-Fall.
  IEEE}, September 2017.

\bibitem{Friese1997multitone}
M.~Friese, ``Multitone signals with low crest factor,'' \emph{Communications,
  IEEE Transactions on}, vol.~45, no.~10, pp. 1338--1344, Oct 1997.

\bibitem{bas_realjournal_2017}
C.~U. Bas and et.al., ``{A Real-Time Millimeter-Wave Phased Array MIMO Channel
  Sounder for Dynamic Measurements },'' to be submitted.

\bibitem{bas_2017_microcell}
C.~U. Bas, R.~Wang, S.~Sangodoyin, S.~Hur, K.~Whang, J.~Park, J.~Zhang, and
  A.~F. Molisch, ``{28 GHz Microcell Measurement Campaign for Residential
  Environment },'' in \emph{{2017 IEEE Global Communications Conference
  (GLOBECOM)}}, December 2017.

\bibitem{massey_1951_kolmogorov}
F.~J. Massey~Jr, ``The kolmogorov-smirnov test for goodness of fit,''
  \emph{Journal of the American statistical Association}, vol.~46, no. 253, pp.
  68--78, 1951.

\end{thebibliography}

\end{document}